\documentclass{Interspeech}

\interspeechcameraready 

\usepackage{hyperref}
\hypersetup{
    colorlinks=true,
    citecolor=blue,
    urlcolor=magenta,
}
\usepackage{colortbl}
\usepackage{newtxtext}

\usepackage{cite}
\usepackage{amsmath}
\usepackage{url}
\usepackage{amsfonts}
\usepackage{multirow}
\usepackage{makecell}
\usepackage{booktabs} % toprule, cmidrule, ...
\usepackage{subcaption}
\usepackage{pythonhighlight}
\usepackage[normalem]{ulem}
\useunder{\uline}{\ul}{}

\usepackage{pifont}

\newcommand{\ours}{\textsc{SHEET}}

\title{SHEET: A Multi-purpose Open-source\\Speech Human Evaluation Estimation Toolkit}

\author[affiliation={1}]{Wen-Chin}{Huang}
\author[affiliation={2}]{Erica}{Cooper}
\author[affiliation={3}]{Tomoki}{Toda}

\affiliation{Graduate School of Informatics}{Nagoya University}{Japan}
\affiliation{}{National Institute of Information and Communications Technology}{Japan}
\affiliation{Information Technology Center}{Nagoya University}{Japan}
\email{wen.chinhuang@g.sp.m.is.nagoya-u.ac.jp}

\usepackage{comment}

\begin{document}

\maketitle

\begin{abstract}
    We introduce SHEET, a multi-purpose open-source toolkit designed to accelerate subjective speech quality assessment (SSQA) research. SHEET stands for the Speech Human Evaluation Estimation Toolkit, which focuses on data-driven deep neural network-based models trained to predict human-labeled quality scores of speech samples. SHEET provides comprehensive training and evaluation scripts, multi-dataset and multi-model support, as well as pre-trained models accessible via Torch Hub and HuggingFace Spaces. To demonstrate its capabilities, we re-evaluated SSL-MOS, a speech self-supervised learning (SSL)-based SSQA model widely used in recent scientific papers, on an extensive list of speech SSL models. Experiments were conducted on two representative SSQA datasets named BVCC and NISQA, and we identified the optimal speech SSL model, whose performance surpassed the original SSL-MOS implementation and was comparable to state-of-the-art methods.
\end{abstract}

\section{Introduction}

Speech quality assessment (SQA) refers to the task of evaluating the quality of speech signals \cite{sqa-2011, speech-evaluation-review-2011, speech-evaluation-review-2024}, and is an essential component to various applications, including telecommunications and speech generation tasks, from text-to-speech (TTS), voice conversion (VC) to speech enhancement. The gold standard for evaluating speech signals is to ask human listeners to assign subjective ratings based on perceived quality, using protocols like the mean opinion score (MOS) test. However, since conducting such evaluations can be expensive and time-consuming, objective metrics have been proposed. Among those, some metrics evaluate dimensions like clarity and intelligibility (like the short-time objective intelligibility measure (STOI) \cite{stoi}) or rely on human-defined distances between hand-crafted features (like the Mel cepstral distortion (MCD) \cite{mcd}).

We are particularly interested in the task of developing metrics that are \textbf{optimized directly using human preference data}, where one representative early attempt was the perceptual evaluation of speech quality (PESQ) metric \cite{pesq}. We term such a task \textbf{subjective speech quality assessment} (SSQA). Since such metrics are often data-driven and based on machine learning models, it is no exception for the field of SSQA to benefit from the rapid development of deep neural networks (DNNs) in recent decades \cite{automos, mosnet, dnsmos}. A scientific competition series for promoting SSQA named the VoiceMOS Challenge (VMC) \cite{voicemos2022, voicemos2023, voicemos2024} was founded in 2022, and in that year, it was shown that the best-performing system, UTMOS \cite{utmos}, achieved a high correlation (0.959) with human ratings. Such results led to the adaptation of SSQA models as an objective measure for speech quality in TTS research and nourished increasing interest in SSQA as a critical research area.

\begin{table}[t]
    % \footnotesize
    \centering
    \caption{Comparison of existing open-source speech quality assessment toolkits and \ours{}.}
    \label{tab:comparison}
    
    \begin{tabular}{c | c c c c}
        \toprule
        Toolkit & Inference &  \makecell{Model\\training} & \makecell{Multi-\\model} & \makecell{Multi-\\dataset} \\ 
        \midrule
        \cite{torchaudio-squim, speechbertscore, versa}, etc. & \checkmark & & \checkmark & \\
        \cite{nisqa, ssl-mos, utmos, ramp}, etc. & \checkmark & \checkmark & & \\
        \ours{} & \checkmark & \checkmark & \checkmark & \checkmark \\
        \bottomrule
    \end{tabular}
\end{table}

In the modern era of artificial intelligence research, open-source activities play an important role in promoting and facilitating development. In the field of SSQA, existing open-source toolkits can be categorized into two types, as summarized in Table~\ref{tab:comparison}. The first type (row one) aims to provide an easy-to-use interface to multiple off-the-shelf metrics and pre-trained SQA models \cite{torchaudio-squim, speechbertscore, versa}. However, these toolkits often do not provide training recipes, making it difficult for researchers to develop custom models. The second type (row two) is often reproducible training recipes for their specific scientific papers \cite{nisqa, dnsmos, ssl-mos, utmos, ramp}, typically supporting only a limited range of datasets and models. To summarize, the lack of flexibility of existing toolkits poses challenges for researchers seeking to experiment with different datasets and architectures.

To address these limitations, we introduce \textbf{\ours{}}\footnote{\url{https://github.com/unilight/sheet}}, an open-source toolkit designed to facilitate SSQA research. \ours{} stands for the \textbf{S}peech \textbf{H}uman \textbf{E}valuation \textbf{E}stimation \textbf{T}oolkit, which is designed with the following use cases in mind:
\begin{itemize}
    \item Provide complete training and evaluation scripts for conducting SSQA experiments, making it accessible for newcomers to SSQA research.
    % These scripts are well-documented and designed to be modular, allowing users to customize various aspects of the training and evaluation process. This ensures that researchers can easily adapt the framework to their specific needs and quickly set up their experiments without extensive modifications.
    \item Enable researchers with existing SSQA models to evaluate their models on multiple testing sets using a standardized framework. 
    % Our toolkit supports a variety of benchmark datasets, offering built-in preprocessing and evaluation pipelines. By providing seamless dataset integration, SHEET ensures that researchers can compare their models fairly and efficiently across different datasets without additional effort. Furthermore, we provide tools for analyzing and visualizing model performance, making it easier to interpret results and identify areas for improvement.
    \item Offer pre-trained SSQA models that can be easily accessed through \texttt{torch.hub.load} or HuggingFace.
\end{itemize}

To demonstrate the capabilities of \ours{}, we conducted an extensive experiment using SSL-MOS \cite{ssl-mos}, a representative SSQA model,  to re-evaluate the effectiveness of existing self-supervised learning (SSL) models in predicting human judgments. Experiments were conducted on two datasets: BVCC \cite{bvcc}, which consists of speech samples generated by a total of 187 TTS and VC systems, and NISQA \cite{nisqa}, covering simulated and real-world speech samples in a wide range of distorted conditions (noise, packet-loss, warping, low-bitrate, etc). We identified the optimal SSL model for each dataset, whose result surpassed the original SSL-MOS model and is comparable with the state-of-the-art method. Such results underscore the usefulness of the proposed toolkit and its adaptability to different datasets.

\section{\ours{} design and supported models}

\subsection{Overview}

The structure of \ours{} follows the design of popular speech processing toolkits like Kaldi \cite{kaldi} and ESPnet \cite{espnet}. Specifically, in such a design, there are two core parts: the \textit{library} implements the model architectures, loss computation, and training logic, which are based on Python. On the other hand, a \textit{recipe} is a collection of bash or python scripts that provides easy-to-understand instructions to complete an experiment, from data preprocessing, model training to benchmark evaluation. In practice, each training dataset has one recipe, and there are many configuration files to choose from, each representing a set of hyper-parameters of the model and the training procedure.

\subsection{Supported datasets}

% \ours{} currently provides training recipes for a total of seven training datasets, including BVCC \cite{bvcc}, SOMOS \cite{somos}, SingMOS \cite{singmos}, NISQA \cite{nisqa}, TMHINT-QI \cite{tmhintqi}, Tencent \cite{conferencingspeech2022} and PSTN \cite{pstn}. Evaluation could be done with the twelve supported testing sets, sourcing from the above-mentioned datasets and the previous VMCs. This collection demonstrates a diverse collection of datasets with different properties.
\ours{} currently provides training recipes for a total of seven training datasets and twelve testing sets, demonstrating a diverse collection of datasets with different properties.
The domains span from samples synthesized by speech generation systems like TTS and VC to speech that underwent a variety of distortions, including artificially added and real noise, reverberation, VoIP, transmission, and replay. A total of six languages were covered, and the sampling frequencies also ranged from 8000 Hz to 48000 Hz. Finally, some training datasets provide listener-wise scores, allowing for listener modeling techniques.

Due to space limits, we omit the introduction to each training and testing dataset. Interested readers and users can refer to individual papers, and access the recipes as they are already made publicly available.

\begin{figure}[t]
    \centering
    \includegraphics[width=\linewidth]{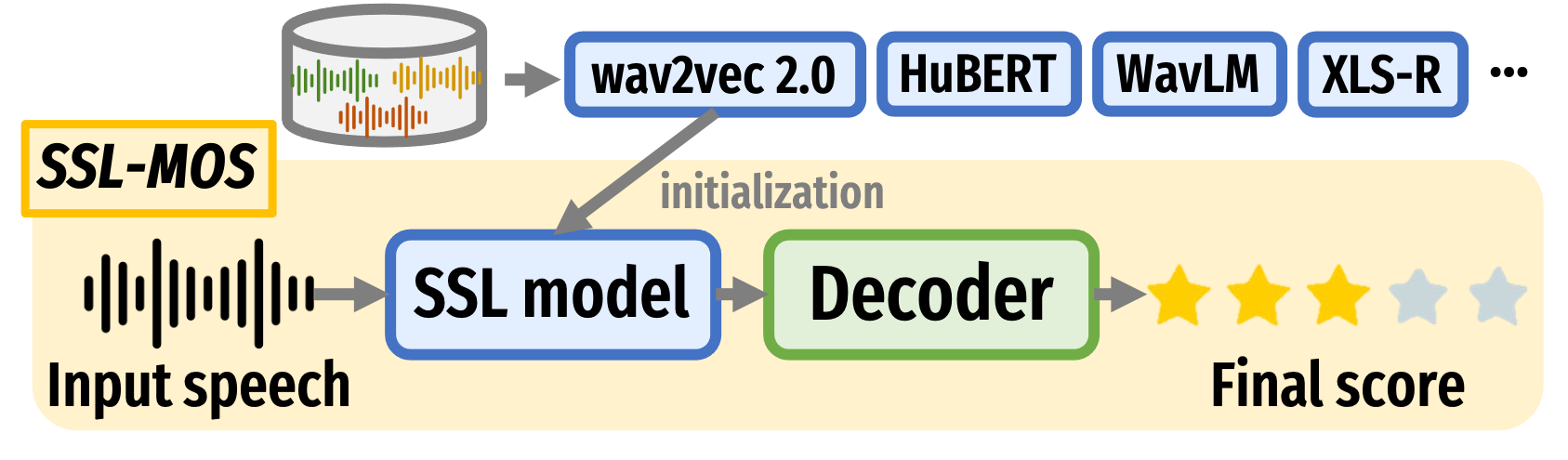}
    \caption{Illustration of SSL-MOS.}
    \label{fig:ssl-mos}
\end{figure}

\subsection{Supported models and features}
\label{ssec:models}

Early DNN-based SSQA models were mainly built upon convolutional and recurrent layers and trained from scratch using a certain SSQA dataset \cite{automos, mosnet, e2e-non-intrusive-2021, dnsmos, nisqa}. As the SSL paradigm revolutionized almost all fields in speech processing \cite{speech-ssl-review}, SSQA was no exception \cite{mos-ssl-2021, ssl-mos}. Thus, all models supported in \ours{} are based on SSL. Below, we introduce one of the most representative SSL-based SSQA models.

\noindent\textbf{SSL-MOS.} SSL-MOS has been widely used as an objective measure of speech quality in scientific papers and even challenges like the Singing Voice Conversion Challenge \cite{svcc2023}, due to its simplicity and strong performance. Figure~\ref{fig:ssl-mos} is an illustration of SSL-MOS. Given an input speech $\vec{x}$, the SSL model $\textsc{SSL}(\cdot)$ outputs a sequence of frame-wise hidden representations which is further sent into a decoder $\textsc{Dec}(\cdot)$ to generate a frame-wise score sequence. The training objective $\mathcal{L}$ is an L1 loss between the ground truth score $y$ and the predicted score $\hat{y}$ obtained by applying time pooling over the frame-wise score sequence:
\begin{equation}
    \mathcal{L} = \Vert y - \hat{y} \Vert_1 = \Vert y - \text{TimePooling}(\textsc{Dec}(\textsc{SSL}(\vec{x}))) \Vert_1.
\end{equation}
During training, $\textsc{SSL}(\cdot)$ and $\textsc{Dec}(\cdot)$ are jointly optimized. During inference, the process to obtain the predicted score $\hat{y}$ is the same as that during training.

\noindent\textbf{Other features.} \ours{} implements features and components that were reported useful in recent SSQA papers, including:
\begin{itemize}
    \item Range clipping, repetitive padding, and the clipped L1 loss, as proposed in \cite{mbnet}.
    \item Listener modeling, as proposed in \cite{mbnet, ldnet}.
    \item Contrastive loss, as proposed in \cite{utmos}.
    \item Multiple dataset training, as proposed in \cite{alignnet}.
    \item Retrieval augmentation based on k-nearest neighbors, as proposed in \cite{ramp}\footnote{We also attempted to reproduce the full RAMP model \cite{ramp}, which is the current state-of-the-art on BVCC and the top system in VMC 2023 and 2024. However, the official implementation did not release training scripts, and as of the time of the submission, we could not fully reproduce the results as shown in the paper.}.
\end{itemize}

\subsection{Training recipe}

A typical training recipe in \ours{} involves the following stages.

\noindent\textbf{Downloading stage.} We provide automatic downloading scripts or manual instructions to obtain the dataset. To ensure reproducibility, we also provide pre-trained model checkpoints for each dataset, which are hosted on HuggingFace.

\noindent\textbf{Data preparation stage.} We provide scripts to integrate all necessary information for training, inference, and evaluation into files in \texttt{.csv} format. Information includes the audio \texttt{.wav} path, sample ID, system ID (if available), and human score.

\noindent\textbf{Model training stage.} Given a configuration file (in \texttt{yaml} format, the SSQA model training takes place in this stage. We provide convenient features, including automatic early stopping based on a pre-defined criterion on the validation set, and inference result visualization for inspection during training.

\noindent\textbf{Evaluation stage.} We provide an integrated script to evaluate the trained model on all the supported test sets.

\subsection{Inference usages}

\ours{} adopts the \texttt{torch.hub} function to support an easy deployment of our provided pre-trained model in Python. The following example code demonstrates how to estimate the quality score given an audio sample path.
\begin{python}
import torch

# load model.
# <anonymized> will be updated upon acceptance
predictor = torch.hub.load(
    "<anonymized>/sheet:v0.1.0",
    "default"
)
# prediction. example output: 3.6066928
score = predictor.predict(
    wav_path="/path/to/wav/file.wav"
)
\end{python}
For researchers who wish to benchmark the SSQA model they developed, \ours{} provides instructions to download the testing sets. After predictions are generated, they may utilize the provided calculation script to assess the performance of the SSQA model. 

\section{Experiments}

To demonstrate the capabilities of \ours{}, in this section, we conduct an extensive experiment to re-evaluate the effectiveness of existing SSL models in the SSL-MOS framework.

\subsection{Experimental setting}

\subsubsection{Dataset}

The following two datasets are used in the experiments.
\begin{itemize}
    \item The \textbf{BVCC} dataset \cite{bvcc} was used in the main track of VMC 2022. It contains English speech samples in 16 kHz and their MOS ratings from 187 different TTS and VC systems, which mainly come from past years of the Blizzard Challenges (BC) and Voice Conversion Challenges (VCC), as well as published samples from ESPnet-TTS \cite{espnet-tts}. Each sample was rated by 8 distinct listeners.
    
    \item The \textbf{NISQA} dataset contains several subsets, each containing speech samples with a sampling frequency of 48 kHz. The \texttt{NISQA TRAIN SIM} and \texttt{NISQA VAL SIM} sets contained English speech samples with simulated distortions and background noises, while the \texttt{NISQA TRAIN LIVE} and \texttt{NISQA VAL LIVE} sets contained English live Skype and phone recordings, with real distortions created during recording. Each sample in these four sets was rated by five listeners. We combined the \texttt{NISQA TRAIN SIM} and \texttt{NISQA TRAIN LIVE} sets as the training set, and combined the \texttt{NISQA VAL SIM} and \texttt{NISQA VAL LIVE} sets as the validation set, resulting in 11020/2700 samples, respectively. For the test sets, we used the following sets: \textbf{\texttt{P501}} and \textbf{\texttt{FOR}} sets each included 240 English samples with simulated distortions and live VoIP calls where speech samples were played back directly from the laptop. Each sample in these two sets was rated by thirty listeners. The \textbf{\texttt{LIVETALK}} set consists of 232 real German phone call recordings from different backgrounds and distortions. Each sample in these two sets was rated by 24 listeners.
\end{itemize}

\subsubsection{Model and training settings}

We mainly experimented with the SSL-MOS model, and we utilized S3PRL \cite{s3prl}, an open-source toolkit that provides a convenient interface to access a large collection of speech and audio SSL models. Using the supported models in S3PRL, our experiment spanned 22 SSL models. Due to space limits, we refer authors to the S3PRL codebase for details of each SSL model\footnote{\url{https://s3prl.github.io/s3prl/tutorial/upstream_collection.html}}. The output from the last layer of the SSL model was used by default. Since most SSL models take 16 kHz waveforms as input, all input speech samples were resampled to 16 kHz. The training batch size was set to 16, and the SGD optimizer with an initial learning rate of 0.001 and a momentum of 0.9 was used. All training runs were allowed to execute for a maximum of 100,000 steps, and if the 5 best checkpoints had not been updated for 2000 steps, the training automatically halted. For more detailed settings, readers can refer to $\texttt{egs/bvcc/conf/ssl-mos-wav2vec2.yaml}$, which is a representative configuration file.

\subsubsection{Comparing systems}

As the main purpose of \ours{} is to promote open-source, reproducible SSQA research, our experiments focus on evaluating publicly available model checkpoints of existing systems, instead of just reporting numbers in previous papers. Therefore, we chose five models and used the checkpoints from their official implementation codebases as the comparing systems. The \textbf{SSL-MOS} \cite{ssl-mos}, \textbf{UTMOS} \cite{utmos}, and \textbf{RAMP+} were trained with the BVCC training set. SSL-MOS and UTMOS were strong-performing models in VMC 2022. RAMP+ is an improved version of the previous RAMP model \cite{ramp}, which was the top-performing system in VMC 2023\footnote{The technical paper of RAMP+ has not yet been published as of the submission date of this paper, and the model checkpoints of RAMP were not provided. We therefore only reported RAMP+ results.}. The \textbf{NISQA} model \cite{nisqa} was trained on a large collection of datasets including the NISQA training sets, and \textbf{DNSMOS P808} \cite{dnsmos} was trained using an internal crowdsourcing SSQA dataset, not including the NISQA training sets. The choice of the last two models was driven by their common use in evaluating noisy and distorted speech. Note that the last two models are not based on SSL

\subsubsection{Evaluation metrics}

For BVCC, we reported the following two metrics: system-level mean squared error (Sys MSE) and Spearman's rank correlation coefficient (Sys SRCC). For NISQA, following the original paper \cite{nisqa}, we reported the following two metrics: utterance-level MSE (Utt MSE) and utterance-level linear correlation coefficient (Utt LCC). Readers may refer to $\texttt{sheet/evaluation/metrics.py}$ for the implementation of the calculation.

% Please add the following required packages to your document preamble:
% \usepackage{booktabs}
% \usepackage{multirow}
\begin{table*}[t]

% \tiny
\scriptsize
% \footnotesize
% \small

\centering
\caption{Experimental results. For MSE, the smaller the better, and for LCC and SRCC, the larger the better. Boldface and underline indicate the best and second-best score in each column, respectively. $^{\dagger}$: results calculated using official pre-trained model checkpoints.}
\label{tab:giant-table}

\begin{tabular}{@{}cl|cc|cccccccc@{}}
\toprule
\multirow{2}{*}{Model}                         & \multicolumn{1}{c|}{\multirow{2}{*}{SSL model}} & \multicolumn{2}{c|}{BVCC test}  & \multicolumn{2}{c}{NISQA test FOR}         & \multicolumn{2}{c}{NISQA test LIVETALK}    & \multicolumn{2}{c}{NISQA test P501}        & \multicolumn{2}{c}{NISQA average} \\
                                               & \multicolumn{1}{c|}{}                           & Sys MSE        & Sys SRCC       & Utt MSE        & Utt LCC        & Utt MSE        & Utt LCC        & Utt MSE        & Utt LCC        & Utt MSE         & Utt LCC         \\ \midrule
\multicolumn{1}{c|}{\multirow{19}{*}{SSL-MOS}} & CPC                                             & 0.186          & 0.873          & 0.367          & 0.723          & 0.801          & 0.499          & 0.440          & 0.790          & 0.536           & 0.671           \\
\multicolumn{1}{c|}{}                          & APC                                             & 0.194          & 0.845          & 0.355          & 0.737          & 0.381          & 0.761          & 0.457          & 0.817          & 0.398           & 0.772           \\
\multicolumn{1}{c|}{}                          & VQ-APC                                          & 0.199          & 0.841          & 0.402          & 0.691          & 0.419          & 0.728          & 0.508          & 0.789          & 0.443           & 0.736           \\
\multicolumn{1}{c|}{}                          & NPC                                             & 0.321          & 0.802          & 0.720          & 0.524          & 0.516          & 0.673          & 0.778          & 0.603          & 0.671           & 0.600           \\
\multicolumn{1}{c|}{}                          & DeCoAR 2.0                                      & 0.255          & 0.917          & 0.148          & 0.903          & 0.306          & 0.824          & 0.455          & 0.901          & 0.303           & 0.876           \\
\multicolumn{1}{c|}{}                          & wav2vec                                         & 0.200          & 0.870          & 0.261          & 0.824          & 0.344          & 0.841          & 0.416          & 0.865          & 0.340           & 0.843           \\
\multicolumn{1}{c|}{}                          & vq-wav2vec                                      & 0.166          & 0.870          & 0.270          & 0.806          & 0.469          & 0.747          & 0.568          & 0.841          & 0.436           & 0.798           \\
\multicolumn{1}{c|}{}                          & wav2vec 2.0 base                                & 0.146          & 0.928          & 0.128          & 0.918          & 0.444          & 0.806          & 0.363          & 0.928          & 0.312           & 0.884           \\
\multicolumn{1}{c|}{}                          & wav2vec 2.0 large                               & 0.100    & 0.929    & 0.139          & 0.903          & 0.235          & 0.863          & 0.182          & 0.918          & 0.185           & 0.895           \\
\multicolumn{1}{c|}{}                          & HuBERT base                                     & 0.175          & 0.916          & {\ul 0.117}    & 0.919          & 0.432          & 0.777          & 0.249          & 0.918          & 0.266           & 0.871           \\
\multicolumn{1}{c|}{}                          & HuBERT large                                    & 0.109          & {\ul 0.936} & 0.209          & 0.862          & 0.482          & 0.791          & 0.323          & 0.879          & 0.338           & 0.844           \\
\multicolumn{1}{c|}{}                          & data2vec base                                   & 0.131          & 0.905          & 0.202          & 0.860          & 0.280          & 0.832          & 0.436          & 0.908          & 0.306           & 0.867           \\
\multicolumn{1}{c|}{}                          & data2vec large                                  & {\ul 0.098} & 0.919          & 0.300          & 0.888          & 0.224          & 0.884          & 0.608          & 0.891          & 0.377           & 0.888           \\
\multicolumn{1}{c|}{}                          & WavLM base                                      & 0.131          & 0.920          & 0.154          & 0.906          & 0.331          & 0.809          & 0.305          & 0.919          & 0.263           & 0.878           \\
\multicolumn{1}{c|}{}                          & WavLM large                                     & 0.119          & 0.928          & \textbf{0.098} & \textbf{0.933} & {\ul 0.177}    & {\ul 0.912}    & \textbf{0.140} & \textbf{0.945} & \textbf{0.138}  & \textbf{0.930}  \\
\multicolumn{1}{c|}{}                          & UniSpeech-SAT base                              & 0.118          & 0.917          & 0.131          & {\ul 0.923}    & 0.341          & 0.794          & 0.394          & 0.908          & 0.289           & 0.875           \\
\multicolumn{1}{c|}{}                          & UniSpeech-SAT large                             & 0.118          & 0.925          & 0.150          & 0.895          & 0.217          & 0.888          & 0.230          & 0.918          & 0.199           & 0.900           \\
\multicolumn{1}{c|}{}                          & XLSR                                            & 0.129          & 0.916          & 0.146          & 0.900          & 0.219          & 0.886          & 0.232          & 0.916          & 0.199           & 0.901           \\
\multicolumn{1}{c|}{}                          & XLS-R 1b                                        & 0.124          & 0.922          & 0.121          & 0.920          & \textbf{0.170} & \textbf{0.914} & {\ul 0.154}    & {\ul 0.943}    & {\ul 0.148}     & {\ul 0.926}     \\
\multicolumn{1}{l|}{}                          & BYOL-A 2024-dim                                 & 0.496          & 0.720          & 0.437          & 0.673          & 0.647          & 0.519          & 0.471          & 0.813          & 0.518           & 0.668           \\
\multicolumn{1}{l|}{}                          & SSAST frame                                     & 0.132          & 0.913          & 0.353          & 0.797          & 0.564          & 0.633          & 0.430          & 0.856          & 0.449           & 0.762           \\
\multicolumn{1}{l|}{}                          & MAE-ASR frame                                   & 0.162          & 0.921          & 0.123          & 0.921          & 0.322          & 0.812          & 0.284          & 0.926          & 0.243           & 0.886           \\ \midrule
\multicolumn{2}{l|}{SSL-MOS \cite{ssl-mos}$^{\dagger}$}                                                                      & 0.113          & 0.923          & 0.633          & 0.776          & 1.496          & 0.731          & 0.468          & 0.852          & 0.866           & 0.786           \\
\multicolumn{2}{l|}{UTMOS \cite{utmos}$^{\dagger}$}                                                                       & 0.148          & 0.925          & 0.364          & 0.802          & 1.044          & 0.772          & 0.400          & 0.855          & 0.603           & 0.810           \\
\multicolumn{2}{l|}{RAMP+ $^{\dagger}$}                                                                                         & \textbf{0.086} & \textbf{0.939} & 0.525          & 0.800          & 1.238          & 0.772          & 0.882          & 0.885          & 0.882           & 0.819           \\
\multicolumn{2}{l|}{NISQA \cite{nisqa}$^{\dagger}$}                                                                       & 0.978          & 0.713          & 0.207          & 0.875          & 0.395          & 0.771          & 0.328          & 0.900          & 0.310           & 0.849           \\
\multicolumn{2}{l|}{DNSMOS P808 \cite{dnsmos}$^{\dagger}$}                                                                 & 0.724          & 0.773          & 1.295          & 0.534          & 0.819          & 0.713          & 1.368          & 0.671          & 1.161           & 0.639           \\ \bottomrule

\end{tabular}

\end{table*}

\subsection{Re-evaluating SSL models in SSL-MOS}

Table~\ref{tab:giant-table} shows the experimental results over the 22 SSL models we investigated. On the BVCC test set, the data2vec large model \cite{data2vec} and the HuBERT large model achieved the best Sys MSE and Sys SRCC scores, respectively.
On the NISQA dataset, on average, the WavLM large model \cite{wavlm} and the XLS-R 1b model \cite{xls-r} achieved the best and second best scores on the Sys MSE and Sys SRCC metrics, respectively. The fact that the optimal SSL model for NISQA is different from that of BVCC demonstrates that SSL-MOS is sensitive to the choice of the SSL model.

We also observed some other interesting results. On the BVCC test sets, large SSL models perform better than their base variants. Such a tendency also holds on the NISQA test sets, except for the HuBERT and data2vec model. On the NISQA test LIVETALK set, XLS-R 1b was the best among all SSL models, which is likely because XLS-R 1b was trained in 128 languages, while many other SSL models were trained using English data only. This highlights the importance of pre-training the SSL model with datasets in multiple languages in multi-lingual SSQA. Finally, we included results from three SSL models trained using not only speech but also general audio. Although we expected such a model can learn better representations, especially for NISQA which consists of mostly distorted speech, none of them outperformed other SSL models on both BVCC and NISQA datasets. Nonetheless, we still remain optimistic on this direction.

\subsection{Results with comparing models}

We then compare the best models in the previous section to existing publicly available models. On the BVCC testing set, the SSL-MOS models with HuBERT large and data2vec large outperformed not only NISQA and DNSMOS P808, which were not trained on the BVCC training set but also SSL-MOS and UTMOS, which used the wav2vec 2.0 base model \cite{wav2vec2}. This result could be because of our hyperparameter choices, as well as the use of the features described in Section~\ref{ssec:models}. However, our best results are comparable but still fall behind RAMP+, the state-of-the-art method on the BVCC test set. Reproducing RAMP+ will be an important future work.

On the NISQA testing sets, the SSL-MOS models with WavLM large and XLS-R 1b surpassed all five comparing systems, setting new strong baseline results on the NISQA testing sets for future research. Notably, the results of the NISQA model and DNSMOS fell behind those of many SSL-MOS models explored in the previous section. This again highlights the importance of using SSL models in SSQA.

\section{Conclusions}

In this work, we introduced \ours{}, an open-source toolkit designed to facilitate research in SSQA. By addressing the limitations of existing toolkits, \ours{} provides a standardized and flexible framework for conducting SSQA experiments, evaluating existing models across multiple datasets, or simply employing the provided pre-trained models. To demonstrate its effectiveness, we conducted extensive experiments using SSL-MOS and re-evaluated SSL models on the BVCC and NISQA datasets. Using \ours{}, we were able to easily identify optimal SSL models for different datasets, achieving performance comparable to or better than comparing methods. This highlights the versatility and practical impact of \ours{} in advancing SSQA research.

\ours{} is an ongoing effort, and we aim to continually maintain the toolkit by implementing emerging state-of-the-art methods. Moving forward, we also plan to include methods to evaluate other dimensions in speech, including speaker similarity prediction \cite{voxsim} and descriptive evaluation using natural language \cite{pam, ALLD}.

We also noticed that in Table~\ref{tab:giant-table}, when models trained solely on BVCC (including SSL-MOS, UTMOS, and RAMP+) were tested on NISQA, the performance was worse than that of the NISQA model, and the NISQA model was not even using SSL. This highlights the insufficient out-of-domain generalization ability of current SSQA models. In the search for a general-purpose SSQA model, investigating and improving the out-of-domain generalization ability of SSQA models will be another important future direction.

\section{Acknowledgements}
This work was partly supported by JSPS KAKENHI Grant Number 25K00143 and JST AIP Acceleration Research JPMJCR25U5, Japan.

\bibliographystyle{IEEEtran}
\bibliography{mybib}

\end{document}